\documentclass[superscriptaddress,aps,preprintnumbers,amsmath,amssymb,nofootinbib]{revtex4}
\usepackage{graphicx}
\usepackage{epstopdf}
\usepackage{dcolumn}
\usepackage{bm}
\usepackage{hyperref}
\usepackage{cancel}

\def\a{\alpha}
\def\b{\beta}

\def\k{\kappa}

\def\D{\Delta}

\def\beq{\begin{eqnarray}}
\def\eeq{\end{eqnarray}}

\newcommand{\vev}[1]{ \left\langle {#1} \right\rangle }

\def\slashchar#1{\setbox0=\hbox{$#1$} 
\dimen0=\wd0 
\setbox1=\hbox{/} \dimen1=\wd1 
\ifdim\dimen0>\dimen1 
\rlap{\hbox to \dimen0{\hfil/\hfil}} 
#1 
\else 
\rlap{\hbox to \dimen1{\hfil$#1$\hfil}} 
/ 
\fi}

\begin{document}
\

\title{
A Model of Visible QCD Axion
}

\author{Hajime Fukuda}
\affiliation{Kavli IPMU (WPI), UTIAS, the University of Tokyo, Kashiwa, 277-8583, Japan}
\author{Keisuke Harigaya}
\affiliation{ICRR, the University of Tokyo, Kashiwa, 277-8582, Japan}
\affiliation{Kavli IPMU (WPI), UTIAS, the University of Tokyo, Kashiwa, 277-8583, Japan}
\author{Masahiro Ibe}
\affiliation{ICRR, the University of Tokyo, Kashiwa, 277-8582, Japan}
\affiliation{Kavli IPMU (WPI), UTIAS, the University of Tokyo, Kashiwa, 277-8583, Japan}
\author{Tsutomu T.~Yanagida}
\affiliation{Kavli IPMU (WPI), UTIAS, the University of Tokyo, Kashiwa, 277-8583, Japan}

\begin{abstract}
We pursue a class of visible axion models where the axion mass is enhanced 
by strong dynamics in a mirrored copy of the Standard Model
in the line of the idea put forward by Rubakov.
In particular, we examine the consistency of the models 
with laboratory, astrophysical, and cosmological constraints.
As a result, viable parameter regions are found, where the 
mass of the axion is of $O(100)$\,MeV or above while the Peccei-Quinn breaking 
scale is at around $10^{3{\mbox-}5}$\,GeV.
\end{abstract}

\date{\today}
\maketitle
\preprint{IPMU15-0050}
\section{Introduction}
It is well known that the strong interaction conserves $CP$-symmetry very well:
$CP$-violating processes in the Standard Model so far observed
can be explained by the phase of the CKM matrix.
The $CP$-conserving nature of the strong interaction is, however, quite puzzling
since QCD possesses an intrinsic $CP$-violating parameter, the $\theta$-angle.
In fact,  the effective $\theta$-angle,  $\theta_{\rm eff}= \theta+\arg\det Y_u+\arg\det Y_d$,
which sets the magnitude of $CP$-violation in QCD, is constrained to be very small,
$\theta_{\rm eff} \lesssim 10^{-10}$, from the null observation
of the neutron electric dipole moment\,\cite{Baluni:1978rf,Crewther:1979pi,Shifman:1978bx},
$|d_n| < 2.9\times 10^{-26}e\,$cm ($90$\% CL)\,\cite{Baker:2006ts}.
Here, $Y_{u,d}$ denote the up-type and down-type Yukawa matrices, respectively.
By remembering that the phase of the CKM matrix is of $O(1)$, 
the above constraint amounts to unnatural cancellation between the intrinsic $\theta$-angle
and the $O(1)$ phase of the Yukawa matrices.

The most attractive solution to this strong $CP$-problem is based on the Peccei-Quinn (PQ) 
symmetry\,\cite{Peccei:1977hh}.
There, the $U(1)$ PQ-symmetry is an almost exact symmetry but broken  by the axial anomaly of QCD.
After spontaneous breaking of the PQ-symmetry, the associated Nambu-Goldstone boson, the axion,
obtains a non-vanishing potential due to the axial anomaly.
Eventually, the effective $\theta$-angle is cancelled by the vacuum expectation value (VEV) of the axion 
at the minimum of the axion potential.

The original realization of the axion\,\cite{Weinberg:1977ma,Wilczek:1977pj}, however, 
has been excluded experimentally.
There, the axion field is embedded in Higgs doublets
and  the decay constant of the axion, $f_a$, is tied to the electroweak breaking scale, $v_{EW}$, 
i.e. $f_a \simeq v_{EW}$. 
Then the mass of the axion field is roughly given by,
\begin{eqnarray}
m_a \sim \frac{f_\pi m_\pi}{f_a} =O(100)\, {\rm keV}\ . 
\end{eqnarray}
Here $f_\pi$ and $m_\pi$ denote the decay constant and the mass of the neutral pion, $f_\pi\simeq 93$\,MeV and
$m_{\pi}\simeq 135$\,MeV.
Such a light axion with $f_a\simeq v_{EW}$  has been extensively searched for  
via the decay of mesons and quarkonia, which ends up with
a lower limit on the decay constant, $f_a \gtrsim 10$\,TeV\,(see e.g. \cite{Agashe:2014kda}).

Laboratory constraints can be evaded if the PQ-symmetry breaking scale is 
separated from $v_{EW}$ and at a scale much higher than $v_{EW}$.
For such a large decay constant, however, the axion mass becomes very small and has trouble with 
astrophysics.
Eventually, the lower limit of the decay constant is pushed up to $f_a \gtrsim 10^{9-10}$\,GeV (see e.g. \cite{Raffelt:2006cw}).
Based on these observations, two classes of  models of the invisible axion have been proposed, 
often called KSVZ\,\cite{Kim:1979if,Shifman:1979if} and DSFZ\,\cite{Dine:1981rt,Zhitnitsky:1980tq} axion models,
and their phenomenological and astrophysical/cosmological properties have been extensively studied
(for a review, see e.g. \cite{Kawasaki:2013ae}).

In this paper, we pursue another possibility to evade all the constraints, a heavy axion. 
For that purpose,  we need another source of the axion mass than the QCD dynamics, i.e.
additional breaking of the PQ-symmetry to the axial anomaly of QCD.
Such additional breaking, however, cannot be arbitrary since newly added PQ-breaking terms spoil the 
successful cancellation of the effective $\theta$-angle at the minimunm of the axion potential.
To resolve the dilemma, we follow the idea put forward by Rubakov\,\cite{Rubakov:1997vp}, 
where the QCD dynamics in a
copy of the Standard Model (mirrored Standard Model sector) pushes
up the axion mass\footnote{See \cite{Berezhiani:2000gh,Hook:2014cda} 
for recent works on the heavy axion on the line of the Rubakov idea. However, their models have various unsolved cosmological problems. See also discussion in section\,\ref{sec:Z2}.}. 
There, the effective $\theta$-angles in both sectors are aligned with each other
by a softly broken $Z_2$ exchange symmetry of the Standard Model and its mirrored copy.
Thanks to the alignment, the $\theta$-angles in the two sectors are cancelled simultaneously
at the minimum of the axion potential.
In this study, we carefully examine whether the idea can be realized consistently with 
all the constraints, in particular with cosmological ones, by constructing a concrete model.

The organization of the paper is as follows.
In section\,\ref{sec:dynamics}, we introduce a concrete model of the axion
with the mirrored Standard Model sector.
There, we also summarize laboratory and astrophysical constraints on axion parameters.
In section\,\ref{sec:cosmology1},  we discuss cosmological constraints on the axion.
In section\,\ref{sec:cosmology2},  we discuss cosmological constraints on
particles
in the mirrored sector.
In section\,\ref{sec:Z2}, we discuss how to differentiate mass scales in the mirrored sector 
from those in the Standard Model sector without spoiling the PQ solution to the
strong $CP$-problem.
The final section is devoted to conclusions and discussion.

\section{Mirrored Standard Model and Axion Properties}
\label{sec:dynamics}
Let us first introduce two copies of the Standard Model each of which has a single Higgs doublet.
We name them the Standard Model sector and the mirrored sector, respectively.
In the following, we put primes on parameters and fields in the mirrored sector 
to distinguish them from those in the Standard Model sector.
As mentioned above, we assume that dimensionless parameters in both sectors are equal with each other
due to a $Z_2$ symmetry.
In particular, the effective $\theta$-angles in the two sectors are aligned,
 $\theta_{\rm eff} =  \theta_{\rm eff}'$ at the high energy input scale such as the Planck scale.
The electroweak scale and
the QCD scale in the mirrored sector, on the other hand, 
can be different from those in the Standard Model due to a soft breaking of the $Z_2$ symmetry (see section\,\ref{sec:Z2}).

To realize the PQ-symmetry, we introduce a gauge singlet complex scalar field $\phi$
which couples to vector-like pairs of (anti-)fundamental fermions ($q_L$, $\bar{q}_R$) and  ($q_L'$, $\bar{q}_R'$) 
of $SU(3)_c$ and $SU(3)_{c'}$  via 
\begin{eqnarray}
\label{eq:extraY}
{\cal L} = g \phi q_L\bar{q}_R + g \phi q_L'\bar{q}_R' + \text{h.c.} \ ,
\end{eqnarray}
where $g$ denotes a coupling constant. 
Here, we assume that $\phi$ is even under the $Z_2$ symmetry.
This example is nothing but an extension of the KSVZ axion model\,\cite{Kim:1979if,Shifman:1979if}, 
and the PQ-charges are assigned to be $\phi(+1)$, $q_L\bar{q}_R(-1)$, and $q_L'\bar{q}_R'(-1)$, 
respectively.%
\footnote{%
As is the case of the KSVZ axion model, so-called the domain wall number is one in our model.
Domain walls are unstable and hence our model is free from the domain wall problem.
In viable parameter regions we discuss in the following, domain walls decay much before the axion decouples from the thermal bath.
Axions produced by the decay of domain walls are absorbed into the thermal bath and do not affect the standard cosmology.
}
After $\phi$ obtains a VEV, $\vev{\phi} = f_a/\sqrt{2}$, the argument of $\phi$ becomes an axion field $a$ 
with a decay constant $f_a$.
By integrating out the extra quarks, the axion couples to the Standard model and its mirrored copy via 
\begin{eqnarray}
\label{eq:anomalous}
{\cal L}_{\rm eff} \simeq 
 \frac{1}{32\pi^2}
 \left(\frac{a}{f_a} + \theta_{\rm eff} \right)
(G\tilde G+G'\tilde G')+
\frac{6Q_Y^2}{32\pi^2}\frac{a}{f_a} 
(Y\tilde Y+Y'\tilde Y')\ ,
\end{eqnarray}
where $G^{(\prime)}$ and $Y^{(\prime)}$ denote the field strengths of the $SU(3)_c^{(\prime)}$ and $U(1)_Y^{(\prime)}$ gauge fields.%
\footnote{The gauge fields are normalized so that the gauge coupling constants appear in their kinetic terms.}
As we will discuss at the end of this section, we assume that the extra quarks  mix
with the quarks in the Standard model (and they do so similarly in the mirrored sector).
Thus, they have non-vanishing $U(1)_Y^{(\prime)}$ charges, $Q_Y$.
Let us remind ourselves that the effective $\theta$-angles in the two sectors are aligned with each other even at low energies,
so that they are cancelled at the minimum of the effective potential of the axion.
We will confirm this crucial point in section\,\ref{sec:Z2}.

After chiral symmetry breaking by QCD and QCD$^{\prime}$ dynamics,  
the axion obtains a mass through the mixings to the pions in the two sectors.
In particular, when the dynamical scale of QCD$^{\prime}$ is much larger
than that of QCD, the axion mass is dominated by the contribution of the mirrored sector, i.e.
\begin{eqnarray}
\label{eq:mass}
m_a \simeq \frac{\sqrt {z'}}{1+z'} \frac{f_{\pi'} m_{\pi'}}{f_a}\ ,
\end{eqnarray}
where $z' = m_{u'}/m_{d'}$ denotes the ratio of the up to down quark masses in the mirrored sector.
Due to the $Z_2$ symmetry, it should be very close to the one in the Standard Model, i.e. $z' \simeq z \simeq 0.56$.
In this case,
a heavy axion with a mass of $O(100)$\,MeV can be easily achieved, 
for example, by taking $v_{EW'} \simeq 10^2\times v_{EW} $ and $\Lambda_{\text{QCD}'} \simeq 10^3\times \Lambda_{\rm QCD}$
for $f_a \simeq 10^{4}$\,GeV.

In the rest of this section, let us summarize laboratory and astrophysical constraints on the axion parameters, 
$m_a$ and $f_a$.
The crucial difference of the KSVZ-type axion from the original axion model 
(and the DFSZ-type axion model) is that the axion couples to the Standard Model sector and the mirrored sector 
only through Eq.\,(\ref{eq:anomalous}), and does not couple to Standard Model fermions at the tree-level.
Due to the lack of direct interactions to Standard Model fermions, 
the main decay mode of the axion is the one into two photons through the effective interaction term,
\begin{eqnarray}
\label{eq:effectiveF}
{\cal L}_{\rm eff} \simeq \frac{1}{32\pi^2}
\left(6Q_Y^2- \frac{2(4+z)}{3(1+z)}\right)
\frac{a}{f_a} 
(F\tilde F+F'\tilde F')\ ,
\end{eqnarray}
for $m_a \lesssim 3m_\pi$.
The decay rate of this mode is given by,
\begin{eqnarray}
\Gamma_{a\to2\gamma} =
 \frac{1}{16\pi}\left(6Q_Y^2-\frac{2(4+z)}{3(1+z)}\right)^2
\left(\frac{\a}{4\pi}\right)^2\frac{m_a^3}{f_a^2}\ .
\end{eqnarray}
For $m_a \gtrsim 3 m_\pi$, a mode into three pions becomes dominant,  
and eventually, modes into two gluon jets become dominant for a much heavier axion, $m_a \gg 3m_\pi$. 
The decay rates should be compared with axion models with direct fermion couplings,
where decay modes are dominated by the modes into electrons and muons 
for $m_a > 2 m_e$ and $m_a > 2m_\mu$, respectively.

\begin{figure}
 \begin{center}
  \begin{minipage}{0.4\linewidth}
   \begin{center}
    \includegraphics[width=.95\linewidth]{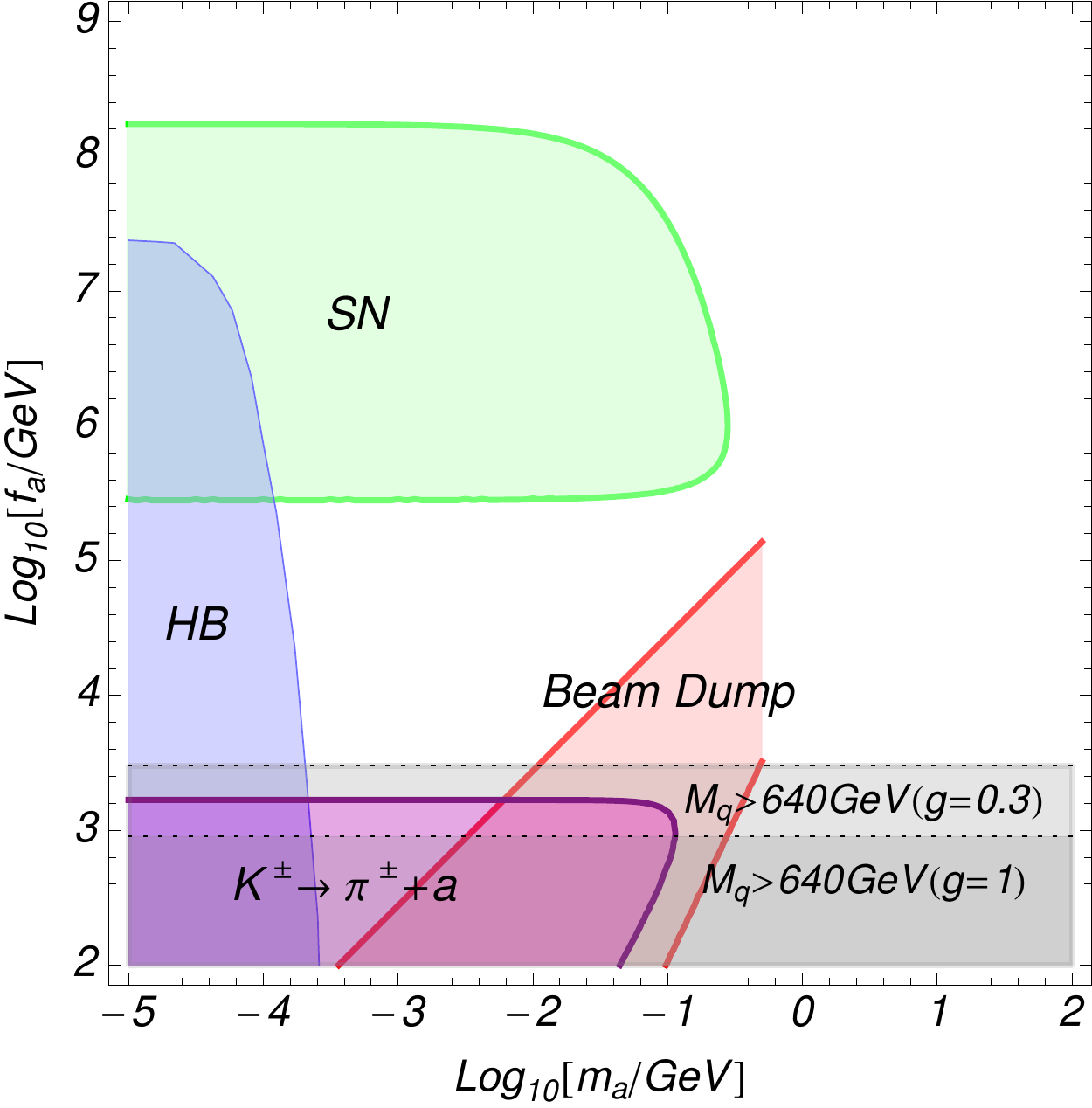}
   \end{center}
  \end{minipage}
 \end{center}
 \caption{
 Constraints on the axion parameters.
The green (light) shaded region labeled by ``SN" denotes the constraint from the supernova neutrino burst duration.
The blue (light) shaded region labeled by ``HB" denotes the constraint from the lifetime of the horizontal branch stars.
The purple (dark) shaded region labeled by ``$K^\pm\to \pi^\pm + a$" denotes the constraint from 
the Kaon decay.
The red shaded region shows the constraint from the proton beam dump experiment CHARM.
Two horizontal lines show the constraint from the extra quark search assuming the Yukawa coupling constant 
in Eq.\,(\ref{eq:extraY}) to be $g = 1$ and $g = 0.3$, respectively.
 }
 \label{fig:axion}
\end{figure}

Accordingly, the laboratory constraints on the axion of this type is quite different from those on models with fermion couplings
(see e.g. \cite{Essig:2010gu} for a compilation of the constraints on the axion-like particle with fermion couplings).
For $m_a \lesssim 0.1$\,GeV, the most stringent constraint comes from the 
 ${\rm Br}[K^\pm\to \pi^\pm + {\rm nothing}]\lesssim 7.3\times 10^{-11}$ at 90\,\% CL\,\cite{Artamonov:2009sz}.
By remembering that the decay of the Kaon into the axion is caused by the $\pi^0-a$ mixing, 
\begin{eqnarray}
{\rm Br}[K^\pm\to \pi^\pm + a\, (\to {\rm invisible})]\simeq \varepsilon_{\pi^0\mbox{-}a}^2 
{\rm Br}[K^\pm\to \pi^\pm + \pi^0]\ , \quad \varepsilon_{\pi^0\mbox{-}a} \simeq \frac{f_\pi(z-1)}{f_a(z+1)}\ ,
\end{eqnarray}
we obtain a constraint, $f_a\gtrsim$ a few TeV for $m_a\lesssim 0.1$\,GeV (see Fig.\,\ref{fig:axion}).%
\footnote{In the figure, we approximate that the size of the E949 detector is about $5$\,m, 
and we require the axion to travel longer than $5$\,m before it decays to contribute to ${\rm Br}[K^\pm\to\pi^\pm+{\rm invisible}]$,
although the lower limit on $f_a$ does not depend on the precise size of the detector significantly.
}
This should be contrasted to axion models with fermion couplings where the dominant contribution 
to the Kaon decay comes from the one-loop Penguin diagrams which leads to a tighter limit on $f_a$,
$f_a \gtrsim O(10)$\,TeV\,\cite{Essig:2010gu}.
It should be also noted that the axion parameters are not constrained  by rare decay of 
quarkonia and $B$-mesons into the axion due to the lack of fermion couplings.%
\footnote{In the decay of quarkonia and $B$-mesons, the axion appears in the final state
through the mixing to $\pi^0$, and hence,
branching ratios into the axion are highly suppressed.}

The axion parameters are also constrained by beam dump experiments.
Again, however, the constraints are much weaker than the case of axion models with fermion couplings.
The most stringent constraint comes from the proton beam dump experiment CHARM at CERN\,\cite{Bergsma:1985qz}.
In Fig.\,\ref{fig:axion}, we translate the constraint in \cite{Bergsma:1985qz}
onto  the KSVZ-type axion model (the red shaded region).
Here, we exclude the parameters which predict at least three events of the axion decay
within the decay region ranging in distance from $445$\,m to $480$\,m 
from the beam dump target.%
\footnote{In our analysis, we assume that the efficiency of the axion signal is independent 
of the mass of the axion and set it to be $0.5$.
We check that our criterion fairly reproduces the constraint at $90$\% CL in \cite{Bergsma:1985qz}
when we apply it to axion models with fermion couplings\,\cite{Bergsma:1985qz}.
}
It should be noted that the constraints are not applicable for $m_a \gtrsim 3m_\pi$, since
the axion decays immediately after it is produced.

In the figure, we also show astrophysical constraints on the axion parameters.
There, the region labeled by ``HB" denotes 
the parameter space in which the lifetime 
of horizontal branch (HB) stars is shorten by the axion production via the 
Primakoff process\,\cite{Raffelt:2006cw}.%
\footnote{We extract the excluded region from \cite{Cadamuro:2011fd}.}
The region labeled by ``SN" denotes the parameter space which 
reduces the
SN 1987A neutrino burst duration.
In the figure, we follow the discussion in\,\cite{Essig:2010gu}, and in particular,  
we allow the parameters with which the mean free path of the axion 
is much less than the supernova core size of $10$\,km.
As the figure shows, astrophysical constraints allow the axion with
a mass above $0.1$\,MeV for $f_a \simeq 10^{4\mbox{-}5}$\,GeV.

For completeness, we also show the constraint on the axion parameters from the 
search for the extra quarks $(q_L,\bar{q}_R)$.
For $f_a\simeq O(1)$\,TeV, the extra quarks obtain their masses from 
Eq.\,(\ref{eq:extraY}), and hence, they are within the reach of collider experiments.
In fact, the production cross section of the extra quarks is much larger than $O(1)$\,fb
at the $8$\,TeV Large Hadron Collider (LHC) experiment when they are lighter than $1$\,TeV. 
In order for the extra quarks to decay immediately, we hereafter assume that extra quarks $q_L$
mix with down-type quarks (and they do so similarly in the mirrored sector) via
\begin{eqnarray}
\label{eq:mixing}
{\cal L} = \xi_i q_L \bar{d}_{Ri}
+
\xi_i' q_L' \bar{d}_{Ri}' + \text{h.c.}\ ,
\end{eqnarray}
where $\xi$s denote small mass mixing parameters and $i$ is the generation index.%
\footnote{The newly added mixing mass parameters do not affect 
the effective $\theta_{\rm eff}$ angles at the tree level since they do not enter the determinant of the mass matrices of quarks.
}
Here, we assume that $q_L^{(\prime)}$ has a vanishing PQ-charge,
so that the mass mixing is consistent with the PQ-symmetry. 
Through these mixings, the extra quarks mainly decay into $H+b$, $Z+b$ and $W+t$,
where we assume that the mixing with the bottom quark is dominant.
To date, the 95\% CL lower limit on the mass of the extra quarks of this type 
is $640$\,GeV set by ATLAS collaboration at the $8$\,TeV running 
with an integrated luminosity $20.3$\,fb$^{-1}$\,\cite{Aad:2015mba}.
In the figure, we show the corresponding exclusion limit on $f_a$ assuming $g=0.3$ and $g=1$, respectively.
This constraint puts the most stringent limit on $f_a$ for a heavy axion, $m_a \gtrsim O(100)$\,MeV.%
\footnote{
The  extra quarks may have a rather long lifetime as long as they do not cause any cosmological problems,
and hence, they can be stable inside  detectors of collider experiments, such as the LHC.
In such cases, the lower limit on $f_a$ gets slightly tighter 
due to the null results of stable exotic hadron searches~\cite{Chatrchyan:2013oca}.
}

\section{Cosmological Constraints on the Axion}
\label{sec:cosmology1}
In the previous section, we have discussed laboratory and astrophysical 
constraints of the KSVZ-type axion model.
We have found that a rather small decay constant $f_a =10^{3{\mbox -}5}$\,GeV 
is consistent with those constraints for $m_a > O(0.1)$\,MeV.
In this section, we discuss whether such parameter regions are consistent with the 
Standard Cosmology.

\begin{figure}
 \begin{center}
  \begin{minipage}{0.4\linewidth}
   \begin{center}
    \includegraphics[width=.95\linewidth]{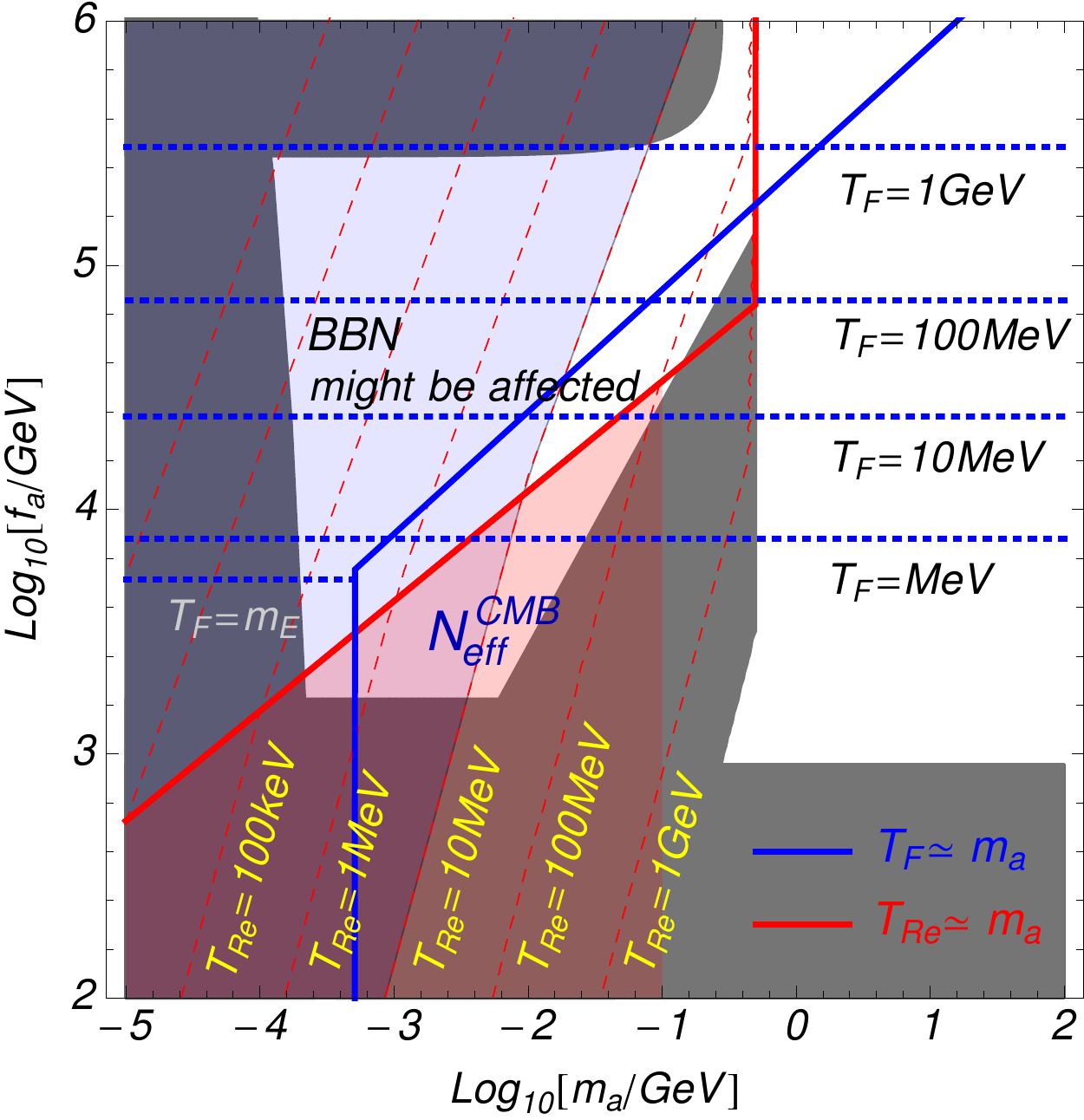}
   \end{center}
  \end{minipage}
 \end{center}
 \caption{
 Constraints on the axion parameters from cosmological arguments.
 The dark shaded region denotes the exclusion limits in Fig.\,\ref{fig:axion}.
The horizontal (blue) dashed lines denote the freeze-out temperature of the Primakoff process.
 The diagonal (red) dashed lines denote the recoupling temperature via the two photon interactions.
The darker (blue) solid line corresponds to the parameters which satisfy $T_{F} \simeq m_a$.
The lighter (red) solid line corresponds to the parameters which satisfy $T_{Re} \simeq m_a$.
The shaded region below the line of $T_{Re} \simeq m_a$ (and labeled by ``$N_{\rm eff}^{\rm CMB}$")
is excluded by the CMB constraints on $N_{\rm eff}$.
The shaded region above the line of $T_{Re} = 10$\,MeV is shown for awareness of the
tension to the BBN (in particular to the D/H abundance).
 }
 \label{fig:cosmology}
\end{figure}

With a rather small decay constant, $f_a =10^{3{\mbox -}5}$\,GeV, 
the axion is kept in thermal equilibrium in the early universe via the effective interactions in Eq.\,(\ref{eq:effectiveF}).
In particular, the axion does not decouple from the thermal bath of the Standard Model
sector until the Primakoff process freezes-out.
In Fig.\,\ref{fig:cosmology}, we show the freeze-out temperature of the Primakoff process $T_F$
given in \,\cite{Cadamuro:2011fd,Millea:2015qra} by horizontal (blue) dashed lines.
The figure shows that $T_F$ is lower than the QCD phase transition temperature,
$T_{\rm QCD} = O(100)$\,MeV in most of the parameter region.
Therefore, the axion could affect the Big-Bang Nucleosynthesis (BBN) and the Cosmic Microwave Background (CMB)
depending on the mass and the lifetime of the axion.
In the figure, a darker (blue) solid line corresponds to the parameters which satisfy $T_{F} \simeq m_a$.
Above this line, the Primakoff process freezes-out when the axion is still relativistic, i.e. $T_F > m_a$.
Below this line, on the other hand, the axion is kept in thermal equilibrium even at $T < m_a$. 
There the axion abundance gets suppressed by a Boltzmann factor until the temperature decreases down to $T_F$.

In the figure,
diagonal (red) dashed lines show
the recoupling temperature  $T_{Re}$
which is defined by
\begin{eqnarray}
\min\left[1,\frac{m_a}{T}\right]\times \Gamma_{a} \simeq 3 H\ ,
\end{eqnarray}
where $H$ denotes the Hubble parameter (see also \cite{Millea:2015qra}).
The darker (red) solid line corresponds to the parameters which satisfy $T_{Re}\simeq m_a$.%
\footnote{In the figure, we find that $T_{Re} > m_a$ is always satisfied 
for $m_a >3 m_\pi$ due to a large decay rate.}
For the parameters above this line, the recoupling temperature is below the axion mass,
i.e. $T_{Re} < m_a$, which means that the axion decays after it becomes non-relativistic.
In the parameter region below this line, on the other hand,  the decay and the inverse decay processes of the axion
freeze in at $T_{Re} > m_a$, which makes the photon in the mirrored sector recouple 
to the thermal bath of the Standard Model.
There, the axion density decreases exponentially by the Boltzmann factor 
when the temperature becomes lower than $m_a$.

Now, let us discuss constraints on the axion parameters from cosmology.
First, let us consider the parameter region which satisfies $T_{Re} > m_a$.
In this region, the mirrored photon is kept in the thermal equilibrium 
with the Standard Model sector until the temperature gets lower than the axion mass.
Thus, the mirrored photon gives sizable contribution to the effective number of relativistic species, $N_{\rm eff}$ 
unless the axion mass is larger than the QCD phase transition.
In addition, both the photon and the mirrored photon are slightly warmed up
by the in-equilibrium decay of the axion.
Putting these contributions together, we find that $N_{\rm eff}$ deviates from the 
Standard Model prediction, $N_{\rm eff}^{\rm SM} = 3.046$\,\cite{Mangano:2005cc} by 
\begin{eqnarray}
{\mit \D} N_{\rm eff} &\simeq&\left( N_{\rm eff}^{\rm SM}  + \frac{8}{7} \right) - N_{\rm eff}^{\rm SM}  \simeq 1.1\ ,  
\quad
(T_{\rm QCD}\gg m_a > T_{\nu\mbox{-}dec})\ ,\\
{\mit \D} N_{\rm eff} &=&\left( N_{\rm eff}^{\rm SM}
\left({11}/{12}\right)^{4/3}
  + \frac{8}{7} \right) - N_{\rm eff}^{\rm SM}  \simeq 0.83\ , \quad
(T_{\nu\mbox{-}dec} > m_a >  T_{e\mbox{-}ann})\ ,
\end{eqnarray}
where $T_{\nu\mbox{-}dec} = O(1)$\,MeV and $T_{e\mbox{-}ann}=O(100)$\,keV denote the neutrino decoupling temperature
and the annihilation temperature of the electron, respectively.
We multiply $8/7$ to the mirrored photon contribution to account for the 
difference of the bosonic and fermionic contributions to $N_{\rm eff}$.
For $m_a < T_{\nu\mbox{-}dec}$, we have also taken into account the relative dilution of the neutrino contribution
to $N_{\rm eff}$  due to the axion decay.%
\footnote{The ${\mit \D}N_{\rm eff}$ is different from the one given in \cite{Millea:2015qra}
because the axion in this model decays into both the photon and the mirrored photon evenly,
and because the mirrored photon contributes to $N_{\rm eff}$.
}
By compared with the constraints on $N_{\rm eff}$ from the CMB observation,
$N_{\rm eff} = 3.15\pm0.23$\,\cite{Planck:2015xua},%
\footnote{If we allow the baryon-to-photon ratio change from the 
best fit value either, the constraint on $N_{\rm eff}$ gets slightly
weaker\,\cite{Planck:2015xua}, although the above deviation, ${\mit \D}N_{\rm eff}\simeq 1$,
has been still excluded even with such weaker constraints.}
we find that the parameter space of $T_{Re} > m_a$ is excluded for $m_a \ll T_{\rm QCD}$.
In Fig.\,\ref{fig:cosmology}, the region labeled by ``$N_{\rm eff}^{\rm CMB}$" denotes the region
excluded by the axion decay contribution to the $N_{\rm eff}$.

Next, let us consider the parameter region which satisfies $T_{Re} < m_a$.
As shown in the figure, most of such parameter space also satisfies 
$T_{F} > m_a$, and hence, the axion in this parameter region decouples 
from the thermal bath when the axion is relativistic and decays after it gets non-relativistic, 
that is, the decay of the axion is of out-of equilibrium.
In addition, contrary to the case of $T_{Re} > m_a$, the mirrored photon does not recouple 
to the thermal bath of the Standard Model sector.
With these two differences, the decay of the axion contributes to $N_{\rm eff}$ differently from the previous case.
To infer the contribution to $N_{\rm eff}$, let us notice the energy density relations at around the 
decay time of the axion, $T \simeq T_{Re}$,
\begin{eqnarray}
\rho_{\gamma + e +(\nu)} &\simeq& \rho_{\gamma + e +(\nu)}(T_{Re})  + \frac{1}{2} \rho_a(T_{Re}) \ ,\\
\rho_{\gamma' } &\simeq& \frac{1}{2}\rho_{a}(T_{Re}) \  . 
\end{eqnarray}
Here, $\rho_a(T_{Re})$ is given by 
\begin{eqnarray}
\rho_a \simeq \frac{\zeta[3]}{\pi^2} \frac{g_{*S}(T_{Re})}{g_{*S}(T_F)} m_a T_{Re}^3\ ,
\end{eqnarray}
where $g_{*S}$ denotes the effective massless degrees of freedom contributing to the entropy density
in the Standard Model.
From these relations, we obtain
\begin{eqnarray}
{\mit \D} N_{\rm eff} &=&\left( N_{\rm eff}^{\rm SM}  +
\frac{8}{7}\frac{g_{*S}(T_{Re})}{2}
\frac{\kappa}{1+\kappa}
\right)
 - N_{\rm eff}^{\rm SM} \ ,  \,
 (T_{\rm QCD}\gg m_a > T_{\nu\mbox{-}dec})\ , \\
{\mit \D} N_{\rm eff} &=&
\left( N_{\rm eff}^{\rm SM}  +
\frac{8}{7}
\frac{g_{*S}(T_{Re})}{2}\kappa
\right)(1+\kappa)^{-1}- N_{\rm eff}^{\rm SM}
\ ,
(T_{\nu\mbox{-}dec} > m_a >  T_{e\mbox{-}ann})\ ,
\end{eqnarray}
where $\kappa$ is defined by,
\begin{eqnarray}
\kappa \simeq\frac{1}{2} \frac{30\zeta[3]}{\pi^4} \frac{1}{g_{*S}(T_F)}\frac{g_{*S}(T_{Re})}{g_*(T_{Re})}\frac{m_a}{T_{Re}}\ .
\end{eqnarray}
Here, $g_*$ denotes the effective massless degrees of freedom contributing to the energy density
in the Standard Model.
By numerical calculation, we find that $\kappa\lesssim 0.1$ in the parameter space for $T_{Re} < m_a$,
and hence, the resultant ${\mit \D}N_{\rm eff}$ is consistent with the constraint from the CMB observation.

It should be noted that the axion decay also affects the baryon-to-photon ratio $\eta$, which 
alters the predictions of the BBN.
In the region of $T_{Re} > m_a$, the baryon-to-photon ratio measured in the CMB observation, 
$\eta_{\rm CMB}$, corresponds to $\eta = (1+\kappa)^{3/4}\eta_{\rm CMB}$ before the axion decay.
By remembering that the primordial D/H abundance which is highly sensitive to $\eta$
is measured precisely, even a slight change of $\eta$ leads to inconsistency between
the BBN prediction and the measurements of the D/H abundance.
To derive precise exclusion limits on the axion parameters, 
however, delicate analysis
involving the evolution of the axion energy density along the BBN is required, 
and it goes beyond this paper.
Here, instead, we lightly shade the region where the decay of the axion could affect the BBN
to note this issue.  

So far, we have implicitly assumed that the electron in the mirrored sector decouples from the thermal bath before the QCD phase transition of the Standard Model sector. 
By remembering that there is a Primakoff process between the axion and the mirrored electron,
the decoupling before the QCD phase transition requires 
either the axion or the mirrored electron must be heavier than $T_{\rm QCD}$.
In these cases, the annihilation of the mirrored electron does not affect the above discussion.

Let us comment on what happens if both the axion and the mirrored electron masses are below
the QCD scale.
In this case the mirrored electron annihilates into the mirrored photon after the QCD phase transition.
Then, for $T_{\rm Re} < m_a$ (i.e.~out-of equilibrium decay), 
the resultant mirrored photon contributes to $N_{\rm eff}$, leading to ${\mit \D}N_{\rm eff} \simeq 2$,
which contradicts with the CMB observations.
For $T_{Re} > m_a$ (i.e. in equilibrium decay), the mirrored photon eventually recouples 
to the thermal bath of the Standard Model sector at the temperature below the QCD scale.
In this case, the resultant mirrored photon from the mirrored electron annihilation can be redistributed between the two sectors.
Such a parameter space, however, has been excluded already by the constraints on $N_{\rm eff}$ as discussed above. 

In summary, we have examined the consistency of the model with cosmology.
As a result, we have found that: 
\begin{itemize}
\item The axion with $T_{Re} > m_a$ and $m_a \lesssim O(100)$\,{MeV}
is excluded by the constraint on $N_{\rm eff}$
of the mirrored photon contribution.
\item The axion with $T_{Re} < m_a$ and $m_a\lesssim O(100)$\,{MeV}
could affect the BBN (the D/H abundance), although delicate analysis is required. 
\item The axion with $m_a > O(100)$\,MeV does not cause cosmological problems.%
\footnote{The freeze-out temperature of the mirrored photon production via
off-shell exchanges of the axion is much higher than the QCD scale even for $f_a \simeq 1$\,TeV.}
\end{itemize}

\section{Cosmological constraints on the Mirrored Sector}
\label{sec:cosmology2}
In this section, let us discuss cosmological constraints on
particles in the mirrored sector.
Most of unstable particles in the mirrored sector decay very fast.
Thus, they  cause no  cosmological problems.
Stable particles, $\gamma'$ $e'$, $\nu'$, $p'$ and $n'$ could, on the other hand,
cause serious cosmological problems unless their abundances are sufficiently suppressed.
As we have already discussed above, for example, the mirrored electron should 
decouple from the Standard Model sector before the QCD phase transition,
since otherwise it increases the mirrored photon abundance.

First, let us discuss the fate of neutrinos in the mirrored sector.
In the Standard Model sector, we assume the seesaw mechanism 
to explain the small neutrino mass\,\cite{seesaw}.
If the same mechanism works in the mirrored sector, the neutrino masses
in the mirrored sector, $m_{\nu'}$, get enhanced by 
\begin{eqnarray}
m_{\nu'} = \frac{v_{ EW'}^2}{v_{ EW}^2} \times m_{\nu}\ .
\end{eqnarray}
As we will discuss in the next section, we mainly consider that $v_{ EW}'/v_{ EW}\gg 1$ 
to make the axion heavy enough, i.e. $m_a\gtrsim O(100)$\,MeV.
Thus, the neutrino masses generated by the seesaw mechanism are much larger than those in the Standard Model sector. 
Eventually, the relic density of the mirrored neutrino exceeds the observed dark matter density in most parameter
region.%
\footnote{Even if the abundance is lower than the observed dark matter density,
there is a hot dark matter constraint, $\sum m_{\nu'} \lesssim 10$--$20$\,eV\,\cite{Viel:2005qj},
which amounts to $v_{EW'} \lesssim 10\times v_{EW}$.}
In order to evade this problem, the seesaw mechanism should not work in the mirrored sector.
This can be achieved by turning off spontaneous breaking of 
the $B-L$ symmetry in the mirrored sector (see discussion in the next section) 
and making the Majorana mass of the right-handed neutrino in the mirrored sector vanish.

Once the seesaw mechanism is turned off in the mirrored sector, neutrinos in 
the mirrored sector obtain the Dirac neutrino mass, 
\begin{eqnarray}
m_{\nu'} \sim \left(\frac{M_R m_{\nu}}{v_{EW}^2} \right)^{1/2} \times v_{EW'}\ ,
\end{eqnarray}
which can be much heavier than the pion in the mirrored sector.
Here, $M_R$ denotes the mass of the right-handed neutrino in the Standard Model sector.
With these large masses, mirrored neutrinos immediately decay into a pair of the 
electron and the pion in the mirrored sector, $\nu'\to e' + \pi'$.
Therefore, neutrinos in the mirrored sector do not cause cosmological problems 
as long as the seesaw mechanism in the mirrored sector is turned off.

Next, let us consider nucleons in the mirrored sector.
Due to their large annihilation cross sections into mirrored pions, 
the abundance of mirrored nucleons is highly suppressed, 
\begin{eqnarray}
\label{eq:proton}
\Omega_{N'}h^2 \sim 10^{-5} \left(\frac{m_{N'}}{\rm TeV} \right)^2 \ .
\end{eqnarray}
One caveat is that the relic mirrored proton becomes dark matter with long-range self-interactions
since they couple to the massless mirrored photon.%
\footnote{%
If $U(1)_Y$ and $U(1)_{Y'}$ gauge bosons have a kinetic mixing with each other, the mirrored proton also has a long-range interaction with charged Standard model particles.
We assume that the kinetic mixing is negligible.
}
The mass density fraction of such dark matter is  constrained roughly below $O(1)$\%\,\cite{McCullough:2013jma}.
Thus, as long as the mirrored proton is lighter than $O(1$--$10)$\,TeV, relic nucleons do not cause
cosmological problems.
It should be noted here that the abundance in Eq.\,(\ref{eq:proton}) assumes no baryon asymmetry
in the mirrored sector.
This assumption is quite natural if we assume that the leptogenesis\,\cite{leptogenesis} 
explains the baryon asymmetry in the Standard Model, since the absence of the seesaw mechanism in the mirrored sector
automatically means the absence of the baryon asymmetry in the mirrored sector.

Finally, let us discuss the fate of pions in the mirrored sector.
Although pions in the Standard Model sector are unstable, 
the charged pions in the mirrored sector are stable since we have assumed that neutrinos are 
heavier than pions in the mirrored sector.
The main annihilation mode of the charged pion is the one into the mirrored photon
with an annihilation cross section,
\begin{eqnarray}
\sigma v \simeq \frac{2\pi \alpha'^2}{m_{\pi'}^2}\ ,
\end{eqnarray}
where $\alpha'$ denotes the fine-structure constant in the mirrored sector.
Accordingly, the relic abundance is roughly given by,
\begin{eqnarray}
\label{eq:pion}
\Omega_{\pi'} h^2 \sim 10^{-4} \left(\frac{m_{\pi'}}{10\,\rm GeV}\right)^2\ .
\end{eqnarray}
This abundance is  sufficiently small as a dark matter component with long-range interactions
as long as $m_{\pi'} \lesssim 100$\,GeV.  
Therefore, mirrored charged pions in this mass range do not cause cosmological problems.

\section{Use of softly broken $Z_2$ symmetry}
\label{sec:Z2}
So far, we have treated the QCD scale and the electroweak scale in the mirrored sector 
as free parameters.
In this section, we discuss how to achieve those mass scales in the mirrored sector by using a 
softly broken $Z_2$ symmetry and show that the crucial condition $\theta_\text{eff} \simeq \theta_\text{eff}'$ is maintained even after the breakdown.
We also discuss how to differentiate the nature of spontaneous breaking of $B-L$ symmetry
in the two sectors.

\begin{figure}
 \begin{center}
  \begin{minipage}{0.4\linewidth}
   \begin{center}
    \includegraphics[width=.95\linewidth]{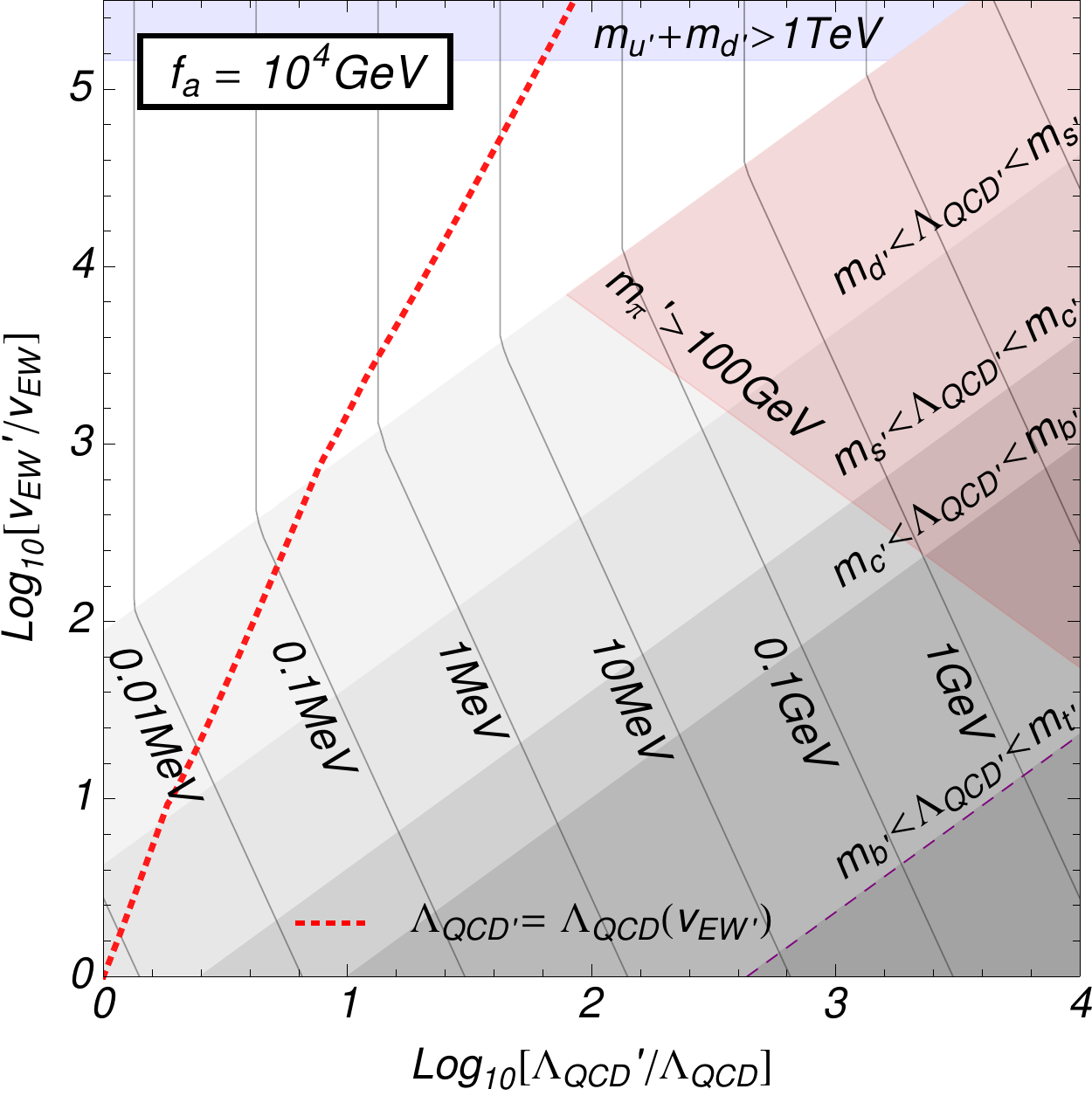}
   \end{center}
  \end{minipage}
 \end{center}
 \caption{The contour plot of the axion mass for $f_a = 10^4$\,GeV.
In the gray shaded regions, 
the relative sizes between the dynamical scale and the quark masses
are different from those in the Standard Model sector.
The red shaded  (right-upper corner) region is excluded since the mirrored pion mass 
is larger than $100$\,GeV.
In the blue shaded (upper horizontal) region, the masses of the mirror quarks exceed $O(1)$\,TeV. Here, we assume that the seesaw mechanism in the mirrored sector is turned off,
so that the charged pion (or the corresponding quarkonum) is stable.
The (red) dashed line corresponds to the $\Lambda_{\text{QCD}'}$ which is increased 
purely by the effect of the larger $v_{EW'}$.
 }
 \label{fig:mass}
\end{figure}

Before
discussing the origin of the scales, however, let us summarize the 
relation between these scales and the axion mass.
In Fig.\,\ref{fig:mass}, we show the contour plot of the axion mass for $f_a = 10^4$\,GeV.
In the figure, we choose $\Lambda_{\rm QCD} \simeq 400$\,MeV.
In the gray shaded regions, 
the relative sizes between the dynamical scale and the quark masses
are different from those in the Standard Model sector.
In our analysis, we assume that the chiral phase transition happens even when 
more than three  quarks in the mirrored sector are lighter than the dynamical scale.
We also assume that the extra quarks in Eq.\,(\ref{eq:extraY}) is heavier 
than the dynamical scale for simplicity.
It should be noted that the electroweak symmetry in the mirrored sector is mainly
broken by the strong ${\rm QCD}^{\prime}$ dynamics in the parameter space with $\Lambda_{\text{QCD}'} \gtrsim v_{EW'}$.
There, the condensation scale of the mirrored Higgs field also becomes $O(\Lambda_{\text{QCD}'})$,
and hence, the nominal parameter $v_{EW'}$ is meaningless.%
\footnote{%
In the figure, we do not show the region with $\Lambda_{\rm QCD'}>f_a$.
Even if we take the potential of $\phi$ such that $\vev{\phi} < \Lambda_{\rm QCD'}$, the condensation of the mirrored extra quarks induces $\vev{\phi}\sim \Lambda_{\rm QCD'}$ through the interaction in Eq.~(\ref{eq:extraY}).
}

The red shaded  (right-upper corner) region is excluded since the mirrored pion mass 
is larger than $100$\,GeV.
In the parameter space with $m_{d'} > \Lambda_{\text{QCD}'}$,
no pion results from 
the chiral symmetry breaking in the mirrored sector.
There, the axion mass is no more given by Eq.\,(\ref{eq:mass}) 
but it scales as $\Lambda_{\text{QCD}'}^2$.
It should be also noted that the hadron picture is no more reliable 
in this parameter space, 
and hence, we need separate discussion on cosmological constraints in the heavy quark picture.
In the figure, we show the rough exclusion limit where the mass of the quarkonium made of $u'$ and $d'$ 
is larger than $1$\,TeV and its relic abundance is expected to exceed 
$O(1)$\% of the total dark matter density.%
\footnote{The exclusion regions by either $m_{\pi'} > 100$\,GeV or $m_{u'}+m_{d'}>1$\,TeV
should be understood as rough estimations.}

Here, let us emphasize that we do not need to increase $v_{EW'}$ from $v_{EW}$ by hand 
to achieve a viable heavy axion. That is, for $m_a > T_{\rm QCD}$, the mirrored electron is not required 
to be heavier than $T_{\rm QCD}$, and hence, there is no requirement to have $v_{EW'} \gg v_{EW}$.
As we have mentioned, however, even if we set the mirrored Higgs mass parameter to be equal
to that of the Standard Model sector, $\Lambda_{\rm QCD'}$ is eventually required to 
be larger than $v_{EW}$ to obtain $m_a > T_{\rm QCD}$, where the Higgs VEV is of $O(\Lambda_{\text{QCD}'})$.
Thus, even if we set $v_{EW'}=v_{EW}$ nominally, the actual electroweak scale becomes much larger
than $v_{EW}$ automatically.

Now, let us discuss how to differentiate the scales of the two sectors by
the soft breaking of $Z_2$ symmetry.
For that purpose, let us introduce a spurion field,
$\sigma(\neq 0)$, 
which changes its sign under the $Z_2$ symmetry.
Here $\sigma$ has a mass dimension one.
With the help of the spurion, it is possible to achieve $m_H^2(\sigma) \neq m_{H'}^{2}(\sigma)$
and allow them to take almost any values.
Concretely, we may choose $Z_2$ invariant parameters, $m_0^2$, $m_1$ and $c$, 
\begin{eqnarray}
m_{H}^2(\sigma) &=& m_0^2 + m_1 \sigma + c\, \sigma^2 \ , \\
m_{H'}^2(\sigma) &=& m_0^2 - m_1 \sigma + c\, \sigma^2 \ ,
\end{eqnarray}
so that $m_{H}^2 \ll m_{H'}^2$.%
\footnote{
One may suspect that this kind of ``fine-tuning'' is problematic. In the low scale theory, it indeed seems unnatural. However, generally speaking, mass scales are what should be generated dynamically in a UV theory.
Since we do not know the UV theory, we allow tuning of mass scales. Otherwise, we must worry about the weak scale itself in the first place, but it is beyond the scope of our paper.}
It should be cautioned here that $\sigma$  cannot be arbitrarily large, since it might appear
any complex phases of parameters in the two sectors suppressed by the reduced 
Planck scale, i.e.~$\sigma/M_{\rm PL}$ with opposite signs.
In particular, the $\theta$-angles in the two sectors may depend on $\sigma$ by
 \begin{eqnarray}
{\cal L} =\frac{\sigma}{M_{\rm PL}}G\tilde G-
\frac{\sigma}{M_{\rm PL}}G'\tilde G'\ ,
\end{eqnarray}
with an $O(1)$ common coefficient.
Therefore, there is an upper limit on the size of the spurion, 
\begin{eqnarray}
\frac{\sigma}{M_{\rm PL}} \lesssim 10^{-12}\ ,
\end{eqnarray}
so that too large $\theta_{\rm eff}$ does not appear in the Standard Model sector at the minimum of the axion potential.

Next, let us discuss how to achieve a larger dynamical scale in the mirrored sector.
As utilized in \cite{Berezhiani:2000gh,Hook:2014cda} to achieve a larger axion mass, 
the dynamical scale of the mirrored sector automatically increases 
by taking $v_{EW'} \gg v_{EW}$ since
quarks decouple
at higher energy scales than the Standard Model sector.
In Fig.\,\ref{fig:mass}, we show  $\Lambda_{\text{QCD}'}$ which is increased 
purely by the larger $v_{EW'}$ as a (red) dashed line.
The figure shows, however, that the axion cannot be heavy enough unless 
$v_{EW'}\gg 10^{7}$\,GeV where the quark mass in the mirrored sector exceeds $O(1)$\,TeV.
Thus, in order to achieve a viable axion mass, $m_a >O(100)$\,MeV, we need to
increase $\Lambda_{\text{QCD}'}$ itself.

With the help of $\sigma$, the larger $\Lambda_{\text{QCD}'}$ can be easily 
achieved by introducing extra scalar quarks 
whose masses again depend on $\sigma$, i.e.
\begin{eqnarray}
{\cal L} = \sum_{i=1}^{N_{\tilde q}} \left(m_{\tilde q}^2\left(\sigma\right) |\tilde q_i|^2 + m_{\tilde q'}^2\left(\sigma\right) |\tilde q'_i|^2 \right)\ ,
\end{eqnarray}
where $N_{\tilde q}$ denotes the number of the extra scalar quarks.%
\footnote{Here, the reason why we introduced ``scalar" quark is that they do not 
contribute to the effective $\theta$-angles, although it may be possible to 
consider extra fermionic colored particles without affecting the effective $\theta$ angles.}
As a simple example, let us choose $N_{\tilde q}$ so that
the beta functions of the $SU(3)$ gauge coupling constant vanish when the mass of the scalar quark is negligible.
In this case, the ratio $\Lambda_{\text{QCD}'}/\Lambda_{\rm QCD}$ is roughly given by $m_{\tilde q'}/m_{\tilde q}$.%
\footnote{The mass $m_{\tilde q'}$ should be at most of the order of  $(\sigma M_{\rm PL})^{1/2} \simeq 10^{12}$\,GeV,
since the size of $\sigma$ is constrained to be $\sigma/M_{\rm PL} \lesssim 10^{-12}$.
Otherwise, $m_{\tilde q} \ll m_{\tilde q'}$ cannot be achieved by fine-tuning.
Accordingly, for $\Lambda_{\text{QCD}'}/\Lambda_{\rm QCD} \simeq 10^{4}$, for example,
the mass of extra scalar quarks in the Standard Model sector is of $O(10^8)$\,GeV or smaller, 
and hence, they are in the thermal bath after inflation if the reheating temperature higher than $10^9$\,GeV
as required by thermal leptogenesis.
In such case, we need separate discussions on how to make these particles unstable. }

By the similar token, we can differentiate the nature of spontaneous breaking of the $B-L$ symmetry in the 
two sectors.
That is, by again assuming that the mass terms of the $B-L$ breaking fields $\Phi_{B-L}^{(\prime)}$ in
the two sectors depend on $\sigma$, we can easily turn on/off spontaneous symmetry breaking 
in the two sectors.
As the size of $\sigma$ is limited from above, so is the $B-L$ breaking scale,
$\vev{\Phi}_{B-L} \lesssim 10^{12}$\,GeV in the Standard Model. 
Fortunately, such $B-L$ breaking scale is high enough to allow thermal leptogenesis in the Standard Model sector.

Before closing this section. let us discuss how largely the $\theta$-angles in the two sectors deviate 
with each other by the soft breaking of the $Z_2$ symmetry.
As we have discussed, the tree-level contributions to the differences of the effective $\theta$-angle 
are controlled by the size of $\sigma$, i.e. $\theta_{\rm eff}^{SM} = O(\sigma/M_{\rm PL})$.
Once we take the weak interactions into account, however, there are another sources of the $CP$-violation,
the CKM, the MNS, and Majorana phases
of the two sectors.
Since
the electroweak scale and the structure of the neutrino masses are differentiated between the two sectors, 
the radiative corrections to the effective $\theta$-angles are different in the two sectors.
Fortunately,
those differences appear at most through $O((\alpha_2/\pi)^2)$ effects
further suppressed by flavor mixings and quark masses\,\cite{Ellis:1978hq,Khriplovich:1985jr}, and hence,
their effects on
the $\theta$-angles are highly suppressed.
Radiative corrections including the mass mixing parameters in Eq.\,(\ref{eq:mixing}) also contributes
to the effective $\theta$-angles.
Such contributions are again suppressed by
$(\alpha_2/\pi)^2$
and quark mixings, 
and hence, the resultant deviation of the angles 
are very small.
Therefore, the uses of the softly broken $Z_2$ symmetry do not spoil the success of the PQ solution
to the strong $CP$-problem.

\section{Conclusions and Discussion}
In this paper, we have pursued a KSVZ-type axion model where the axion mass is enhanced 
by a strong dynamics in the mirrored Standard Model sector.
As we have discussed, the model is consistent with all the constraints when the mass of
the axion is of $O(100)$\,MeV or above even for a relatively low PQ-breaking scale, $10^{3{\mbox-}5}$\,GeV.
We have also noticed that turning off the seesaw mechanism in the mirrored sector solves the two problems simultaneously,
the mirrored neutrino abundance and the too large relic mirrored nucleon mass density.
We have also shown that the mass scales of two sectors can be differentiated systematically by using 
a softly broken $Z_2$ symmetry without spoiling the PQ solution to the strong $CP$-problem.

One unsatisfactory aspect of this model is that the axion is no more the candidate for dark matter.
As an interesting alternative, the neutron in the mirrored sector might be a dark matter candidate for  $m_{n'}\simeq 100$\,TeV.
Here, we assume that $m_{n'}\simeq 100$\,TeV is achieved by a large $\Lambda_{\rm QCD'}$,
which induces the electroweak scale in the mirrored sector at around the similar scale. 
Interestingly, in this parameter region, the mass difference between the proton and the neutron in the mirrored sector 
is dominated by QED' quantum corrections, and hence, the neutron  is automatically lighter
than the proton in the mirrored sector.
Therefore, it can be a good dark matter candidate since it does not have long-range self-interactions.
It should be noted, however, that the charged pion mass in the mirrored sector is expected to be $O(100)$\,GeV
in this parameter range, and hence, this possibility might have a tension with the constraint on the mass 
density fraction of matter with long-range interactions\,\cite{McCullough:2013jma}.
This tension can be easily solved, for example, by assuming that there are only two right-handed neutrinos
in each sector, so that one of the left-handed neutrinos in each sector become 
massless\,\cite{Frampton:2002qc,Harigaya:2012bw}.
With this additional assumption, the charged pion in the mirrored sector decays into a charged leptons and a massless neutrino,
so that it does not contributes to the dark matter density.%
\footnote{We will explore more generic possibilities of dark matter candidates in the mirrored sector 
elsewhere.}

Throughout this paper, we have assumed that the $U(1)$ PQ-symmetry is an almost exact symmetry of the model
broken only by the axial anomaly.
It is generically believed, however, that global symmetries are not respected at least by quantum gravity,
and hence, the PQ-symmetry may be explicitly broken by Planck suppressed operators,
\begin{eqnarray}\label{eq:QG}
{\cal L}_{\cancel{PQ}} = \frac{\k}{(n+4)!M_{\rm PL}^{n}} \left(\phi^{n+4} + \phi^{*n+4}\right)\ ,\quad (n> 0) \ ,
\end{eqnarray}
with $\k = O(1)$.
Such higher dimensional operators leave a non-vanishing effective $\theta$-angle at the minimum of the 
axion potential, 
\begin{eqnarray}
{\mit\Delta}\theta_{\rm eff} \sim \frac{\kappa}{2^{(n+2)/2} (n+3)!}\frac{f_a^{n+2}}{M_{\rm PL}^n m_a^2}\ .
\end{eqnarray}
For dimension five operators ($n=1$), we obtain
\begin{eqnarray}
{\mit \Delta} \theta_{\rm eff} \sim 10^{-10} \times\k\left(\frac{f_a}{10^{4}\,\rm GeV}\right)^{3}\left(\frac{10\,\rm GeV}{m_a}\right)^2\ ,
\end{eqnarray}
which is consistent with the current upper limit on the effective $\theta$-angle
if $f_a \lesssim O(10^3-10^4)$ for $m_a =O(0.1-10)$\,GeV.
The stability against possible quantum gravity effects is the merit of axion models with a small decay constant and a large axion mass.
It is interesting that a non-vanishing effective $\theta$-angle may be observed in near future.

We note that the small decay constant and the large axion mass is also advantageous
when one tries to understand the PQ symmetry as an accidental one resulting from other exact gauge symmetries~(see \cite{Harigaya:2013vja} and references therein).%
\footnote{For example, the baryon number conservation in the Standard Model is accidentally guaranteed by gauge symmetries of the Standard Model.
}
In invisible axion models, where $f_a > 10^9$\,GeV, one must forbid PQ-breaking operators up to dimension-ten in order not to induce too large deviation of the effective $\theta$-angle.
It is not trivial to obtain such a high quality of the accidental PQ symmetry.
In our model, as we have discussed above, it is enough to forbid PQ symmetry breaking by renormalizable interactions.

In this paper, we did not copy the PQ symmetry and the PQ breaking field $\phi$.
It is also possible that there exist a mirrored PQ symmetry and a mirrored PQ breaking, with couplings
\begin{eqnarray}
{\cal L} = g \phi q_L\bar{q}_R + g \phi' q_L'\bar{q}_R' + \text{h.c.} \ .
\end{eqnarray}
Assuming that the two PQ symmetries are softly broken down to a single PQ symmetry by the interaction
\begin{eqnarray}
{\cal L} = M^2 \phi \phi'^\dag +  \text{h.c.} \ ,
\end{eqnarray}
we obtain an axion model with vanishing effective $\theta$-angles.%
\footnote{%
In order for the PQ solution to work, $M^2$ must be real. This is guaranteed by the $Z_2$ symmetry. 
}
If the breaking scale $M^2$ is sufficiently small, there are two light axions.

Let us comment on how visible the present axion model is.
Due to a small PQ-breaking scale, the axion may be searched for at high intensity low energy collider experiments
(see \cite{Essig:2010gu,Essig:2013lka} for related axion search).%
\footnote{Unlike the models discussed in  \cite{Essig:2010gu,Essig:2013lka}, the axion
in this model mainly decay into hadrons in most parameter space, and hence, 
we need further study.} 
 In particular, the new beam dump experiment at CERN, the SHiP experiment, 
is expected to cover axion parameter regions with a shorter lifetime (and hence a heavier axion)
than the CHARM experiments~\cite{Alekhin:2015oba}.
Another interesting possibility is the direct production of the axion
and the radial component
$s$ at the LHC experiments.
In fact, since they couple to the gluon rather strongly, they have sizable production cross sections.%
\footnote{For example, for $f_a \simeq 2$\,TeV, the production cross sections of the axion and its scalar partner $s$ 
via the gluon fusion process are $O(100)$\,fb and $O(0.1$--$1)$\,fb for $m_a \simeq 100$\,GeV and 
$m_s \simeq 1$\,TeV, respectively at the $8$--$14$\,TeV LHC.
} 
Once they are produced at the LHC, the axion immediately decays into jets, while 
$s$ decays into a pair of axions which subsequently decay into jets.
When the axion mass is of $O(1)$\,GeV or below, $s$ appears
as a two-jet resonance in the $O(1)$\,TeV region, which is difficult to be distinguished from 
QCD background processes (see e.g. \cite{Chatrchyan:2013qha}). 
When the axion mass if of $O(10$--$100)$\,GeV, on the other hand, $s$ decays into 
two axions which can be distinguished from QCD background processes by 
looking for peaks in the dijet invariant mass distributions made by the decay 
of the axion\,\cite{Khachatryan:2014hpa}.
As we mentioned in section\,\ref{sec:dynamics}, the production of the extra quarks at the LHC experiments
is also an interesting possibility of the present axion model.

It is also possible to search for 
particles in the mirrored sector.
Here, we just list possible detection methods.
At least, mirrored particles couple to Standard Model particles through the PQ breaking field $\phi$.
The PQ breaking field $\phi$ is
produced at the LHC experiments and decays not only into Standard Model particles but also into mirrored particles, which are invisible for detectors in the LHC experiments.
A channel with jet(s) plus missing energy may be useful in searching for the invisible decay.
The possible kinetic mixing between $U(1)_Y$ and $U(1)_{Y'}$ gauge bosons is also interesting.%
Mirrored particles may be produced in the collision of Standard Model particles through the exchange of the gauge bosons.
Thermal relics of stable charged particles in the mirrored sector (see Sec.~\ref{sec:cosmology2}) may be detectable through
CHArged Massive Particle searches.

Before closing this paper, let us also comment that the above discussion can be 
easily extended to the model consisting of two copies of the Standard Model each of which has two Higgs doublets,
so that the PQ-symmetry is realized as in the original Peccei-Quinn-Weinberg-Wilczek (PQWW) axion model (see also
\cite{Berezhiani:2000gh}).
There, the two sectors share a unique PQ-symmetry through quartic couplings between
the two Higgs doublets in the two sectors.
Contrary to the mirrored KSVZ-type model, the axion decay constant $f_a$ is tied
 to the electroweak scale in the mirrored sector, i.e. $f_a \simeq v_{EW}'$.
The quark mass ratios $z^{(\prime)}$ in the two sectors can also differ with each other due to the difference 
of the ratios of the vacuum expectation values of the doublets in each sector, i.e. $\tan\beta
\neq\tan\b^\prime$.
By repeating our discussion, we will find that the axion should be again heavier 
than $O(100)$\,MeV, so that the model is safely consistent with laboratory, astrophysical, and cosmological constraints.
Thus, we will again need to invoke a mechanism which achieves $\Lambda_{\rm QCD'} \gg \Lambda_{\rm QCD}$ 
separately from the size of $v_{EW}'$ (see section \ref{sec:Z2}).

One interesting feature of the mirrored PQWW model is that the $U(1)_{\text{QED}'}$ can be also broken spontaneously depending on the mass parameters and quartic couplings of the Higgs doublets in the two sectors.
In such case, even the ``charged" particles in the mirrored sector can be good dark matter candidates.
We will explore those possibilities elsewhere.

Another interesting phenomenological difference of the PQWW model is that the main decay mode
of the axion is not the one into three pions but into a pair of muons even for $m_a \gtrsim 3m_\pi$.%
\footnote{For $m_a \gtrsim 800$\,MeV, the mode into $\rho+\pi$ opens, and hence, the hadronic modes become
dominant eventually.}
Thus, this type of the axion can be more visible at the future beam dump experiments such as the SHiP experiment.
Furthermore, it is also possible to detect this type of axion by searching for displaced vertices 
inside the detectors of the LHC experiments made by the axion decay\,\cite{Goh:2008xz}.

\vspace{-10pt}
\section*{Acknowledgements} \vspace{-10pt}
The authors thank A.~Kusenko for useful discussion when we start this project.
The authors also thank V.~Rubakov and R.~Peccei for valuable comments on the first version of the draft.
This work is supported by Grant-in-Aid for Scientific Research from the Ministry of Education, Culture, Sports, Science
and Technology (MEXT) in Japan, No. 26104009 and 26287039 (T.T.Y.) and No. 24740151 and 25105011 (M.I.), 
from the Japan Society for the Promotion of Science (JSPS), 
No. 26287039 (M.I.), as well as by the World Premier International Research Center Initiative (WPI), MEXT, Japan. 
The work of K.H. is supported in part by a Research Fellowship for Young Scientists from the Japan Society for the Promotion of Science (JSPS).


\begin{thebibliography}{99} 
\bibitem{Baluni:1978rf} 
  V.~Baluni,
  Phys.\ Rev.\ D {\bf 19}, 2227 (1979).
\bibitem{Crewther:1979pi} 
  R.~J.~Crewther, P.~Di Vecchia, G.~Veneziano and E.~Witten,
  Phys.\ Lett.\ B {\bf 88}, 123 (1979)
  [Erratum-ibid.\ B {\bf 91}, 487 (1980)].
\bibitem{Shifman:1978bx} 
  M.~A.~Shifman, A.~I.~Vainshtein and V.~I.~Zakharov,
  Nucl.\ Phys.\ B {\bf 147}, 385 (1979).
\bibitem{Baker:2006ts} 
  C.~A.~Baker, D.~D.~Doyle, P.~Geltenbort, K.~Green, M.~G.~D.~van der Grinten, P.~G.~Harris, P.~Iaydjiev and S.~N.~Ivanov {\it et al.},
  Phys.\ Rev.\ Lett.\  {\bf 97}, 131801 (2006)
  [hep-ex/0602020].
  
\bibitem{Peccei:1977hh}
  R.~D.~Peccei and H.~R.~Quinn,
  Phys.\ Rev.\ Lett.\  {\bf 38}, 1440 (1977);
  R.~D.~Peccei and H.~R.~Quinn,
  Phys.\ Rev.\ D {\bf 16}, 1791 (1977).
  
\bibitem{Weinberg:1977ma} 
  S.~Weinberg,
  Phys.\ Rev.\ Lett.\  {\bf 40}, 223 (1978).
\bibitem{Wilczek:1977pj} 
  F.~Wilczek,
  Phys.\ Rev.\ Lett.\  {\bf 40}, 279 (1978).


\bibitem{Agashe:2014kda} 
  K.~A.~Olive {\it et al.}  [Particle Data Group Collaboration],
  Chin.\ Phys.\ C {\bf 38}, 090001 (2014).
\bibitem{Raffelt:2006cw} 
  G.~G.~Raffelt,
  Lect.\ Notes Phys.\  {\bf 741}, 51 (2008)
  [hep-ph/0611350].
\bibitem{Kim:1979if} 
  J.~E.~Kim,
  Phys.\ Rev.\ Lett.\  {\bf 43}, 103 (1979).
\bibitem{Shifman:1979if} 
  M.~A.~Shifman, A.~I.~Vainshtein and V.~I.~Zakharov,
  Nucl.\ Phys.\ B {\bf 166}, 493 (1980).
  
\bibitem{Dine:1981rt} 
  M.~Dine, W.~Fischler and M.~Srednicki,
  Phys.\ Lett.\ B {\bf 104}, 199 (1981).
\bibitem{Zhitnitsky:1980tq} 
  A.~R.~Zhitnitsky,
  Sov.\ J.\ Nucl.\ Phys.\  {\bf 31}, 260 (1980)
  [Yad.\ Fiz.\  {\bf 31}, 497 (1980)].
  
\bibitem{Kawasaki:2013ae} 
  M.~Kawasaki and K.~Nakayama,
  Ann.\ Rev.\ Nucl.\ Part.\ Sci.\  {\bf 63}, 69 (2013)
  [arXiv:1301.1123 [hep-ph]].
\bibitem{Rubakov:1997vp} 
  V.~A.~Rubakov,
  JETP Lett.\  {\bf 65}, 621 (1997)
  [hep-ph/9703409].
  
\bibitem{Berezhiani:2000gh} 
  Z.~Berezhiani, L.~Gianfagna and M.~Giannotti,
  Phys.\ Lett.\ B {\bf 500}, 286 (2001)
  [hep-ph/0009290].
\bibitem{Hook:2014cda} 
  A.~Hook,
  arXiv:1411.3325 [hep-ph].
  
\bibitem{Essig:2010gu} 
  R.~Essig, R.~Harnik, J.~Kaplan and N.~Toro,
  Phys.\ Rev.\ D {\bf 82}, 113008 (2010)
  [arXiv:1008.0636 [hep-ph]].
  
\bibitem{Artamonov:2009sz} 
  A.~V.~Artamonov {\it et al.}  [BNL-E949 Collaboration],
  Phys.\ Rev.\ D {\bf 79}, 092004 (2009)
  [arXiv:0903.0030 [hep-ex]].
\bibitem{Bergsma:1985qz} 
  F.~Bergsma {\it et al.}  [CHARM Collaboration],
  Phys.\ Lett.\ B {\bf 157}, 458 (1985).
  
  
  
\bibitem{Cadamuro:2011fd} 
  D.~Cadamuro and J.~Redondo,
  JCAP {\bf 1202}, 032 (2012)
  [arXiv:1110.2895 [hep-ph]].
  
\bibitem{Aad:2015mba} 
  G.~Aad {\it et al.}  [ATLAS Collaboration],
  arXiv:1503.05425 [hep-ex].
\bibitem{Chatrchyan:2013oca} 
  S.~Chatrchyan {\it et al.}  [CMS Collaboration],
  JHEP {\bf 1307}, 122 (2013)
  [arXiv:1305.0491 [hep-ex]].
\bibitem{Millea:2015qra} 
  M.~Millea, L.~Knox and B.~Fields,
  arXiv:1501.04097 [astro-ph.CO].
\bibitem{Mangano:2005cc} 
  G.~Mangano, G.~Miele, S.~Pastor, T.~Pinto, O.~Pisanti and P.~D.~Serpico,
  Nucl.\ Phys.\ B {\bf 729}, 221 (2005)
  [hep-ph/0506164].
  
\bibitem{Planck:2015xua} 
  P.~A.~R.~Ade {\it et al.}  [Planck Collaboration],
  arXiv:1502.01589 [astro-ph.CO].
  
\bibitem{seesaw}
  T. ~Yanagida,
  Conf. Proc. {\bf C7902131}, p.95 (1979);
  M. Gell- Mann, P. Ramond and R. Slansky,
  Conf. Proc. {\bf C790927}, p.315 (1979);
  S.L. Glashow,
  in Quarks and Leptons, Carg\`{e}se 1979,
  eds. M. L\'{e}vy, et al., (Plenum 1980 New York), p. 707.
  See also 
  P.~Minkowski,
  Phys.\ Lett.\  {\bf B67}, 421 (1977).
  
\bibitem{Viel:2005qj} 
  M.~Viel, J.~Lesgourgues, M.~G.~Haehnelt, S.~Matarrese and A.~Riotto,
  Phys.\ Rev.\ D {\bf 71}, 063534 (2005)
  [astro-ph/0501562].

\bibitem{leptogenesis}
  M.~Fukugita and T.~Yanagida,
  Phys.~Lett.~{\bf B174} (1986) 45; 
  For  reviews,
  W.~Buchmuller, P.~Di Bari and M.~Plumacher,
  Annals Phys.\  {\bf 315}, 305 (2005)
  [hep-ph/0401240];
  W.~Buchmuller, R.~D.~Peccei and T.~Yanagida,
  Ann.\ Rev.\ Nucl.\ Part.\ Sci.\  {\bf 55}, 311 (2005)
  [arXiv:hep-ph/0502169];  
  S.~Davidson, E.~Nardi and Y.~Nir,
  Phys.\ Rept.\ \ {\bf 466}, 105  (2008)
  [arXiv:0802.2962 [hep-ph]].
\bibitem{McCullough:2013jma} 
  M.~McCullough and L.~Randall,
  JCAP {\bf 1310}, 058 (2013)
  [arXiv:1307.4095 [hep-ph]].
\bibitem{Ellis:1978hq} 
  J.~R.~Ellis and M.~K.~Gaillard,
  Nucl.\ Phys.\ B {\bf 150}, 141 (1979).
\bibitem{Khriplovich:1985jr} 
 I.~B.~Khriplovich,
 Phys.\ Lett.\ B {\bf 173}, 193 (1986)
 [Sov.\ J.\ Nucl.\ Phys.\  {\bf 44}, 659 (1986)]
 [Yad.\ Fiz.\  {\bf 44}, 1019 (1986)].
\bibitem{Frampton:2002qc} 
  P.~H.~Frampton, S.~L.~Glashow and T.~Yanagida,
  Phys.\ Lett.\ B {\bf 548}, 119 (2002)
  [hep-ph/0208157].
\bibitem{Harigaya:2012bw} 
  K.~Harigaya, M.~Ibe and T.~T.~Yanagida,
  Phys.\ Rev.\ D {\bf 86}, 013002 (2012)
  [arXiv:1205.2198 [hep-ph]].
 
\bibitem{Harigaya:2013vja} 
  K.~Harigaya, M.~Ibe, K.~Schmitz and T.~T.~Yanagida,
  Phys.\ Rev.\ D {\bf 88}, no. 7, 075022 (2013)
  [arXiv:1308.1227 [hep-ph]].
\bibitem{Essig:2013lka} 
  R.~Essig, J.~A.~Jaros, W.~Wester, P.~H.~Adrian, S.~Andreas, T.~Averett, O.~Baker and B.~Batell {\it et al.},
  arXiv:1311.0029 [hep-ph].
\bibitem{Alekhin:2015oba} 
  S.~Alekhin, W.~Altmannshofer, T.~Asaka, B.~Batell, F.~Bezrukov, K.~Bondarenko, A.~Boyarsky and N.~Craig {\it et al.},
  arXiv:1504.04855 [hep-ph].
\bibitem{Chatrchyan:2013qha} 
  S.~Chatrchyan {\it et al.}  [CMS Collaboration],
  Phys.\ Rev.\ D {\bf 87}, no. 11, 114015 (2013)
  [arXiv:1302.4794 [hep-ex]].
\bibitem{Khachatryan:2014hpa} 
  V.~Khachatryan {\it et al.}  [CMS Collaboration],
  JHEP {\bf 1408}, 173 (2014)
  [arXiv:1405.1994 [hep-ex]].
  
\bibitem{Goh:2008xz} 
  H.~S.~Goh and M.~Ibe,
  JHEP {\bf 0903}, 049 (2009)
  [arXiv:0810.5773 [hep-ph]];
  H.~S.~Goh and M.~Ibe,
  Prog.\ Theor.\ Phys.\ Suppl.\  {\bf 180}, 27 (2010).
\end{thebibliography}
\end{document}